\documentclass{ptephy_v1}

\begin{document}

\newcommand{\beq}{\begin{equation}}
\newcommand{\beqa}{\begin{eqnarray}}
\newcommand{\eeq}{\end{equation}}
\newcommand{\eeqa}{\end{eqnarray}}
\newcommand{\simg}{\gtrsim}
\newcommand{\siml}{\lesssim}
\newcommand{\sol}{M_\odot}
\newcommand{\bm}[1]{\mbox{{\boldmath $#1$}}}

\def\d{{\rm d}}
\def\p{\partial}
\def\w{\wedge}
\def\o{\otimes}
\def\f{\frac}
\def\tr{{\rm tr}}
\def\Half{\frac{1}{2}}
\def\half{{\scriptstyle \frac{1}{2}}}
\def\T{\tilde}
\def\RA{\rightarrow}
\def\N{\nonumber}
\def\n{\nabla}
\def\bb{\bibitem}
\def\BE{\begin{equation}}
\def\EE{\end{equation}}
\def\BEA{\begin{eqnarray}}
\def\EEA{\end{eqnarray}}
\def\L{\label}
\def\zero{{\scriptscriptstyle 0}}

\title{Prompt emission from the counter jet of a short gamma-ray burst}

\author{
\name{\fname{Ryo} \surname{Yamazaki}}{1,\ast}, 
\name{\fname{Kunihito} \surname{Ioka}}{2},
and 
\name{\fname{Takashi} \surname{Nakamura}}{2,3} 
}

\address{
\affil{1}{$^1$Department of Physics and  Mathematics, Aoyama-Gakuin University, Kanagawa 252-5258, Japan}\\
\affil{2}{$^2$Center for Gravitational Physics, Yukawa Institute for Theoretical Physics, Kyoto University, Kyoto 606-8502, Japan}\\
\affil{3}{$^3$Department of Physics, Kyoto University, Kyoto 606-8502, Japan}
\email{ryo@phys.aoyama.ac.jp}
}

\begin{abstract}

The counter jet of a short gamma-ray burst (sGRB) has not yet been observed, while
recent discoveries of gravitational waves (GWs) from a binary neutron star (NS) merger GW170817 
and the associated sGRB~170817A have demonstrated that off-axis sGRB jets are detectable.
We calculate the prompt emission from the counter jet of an sGRB
and show that it is typically 23--26~mag in the optical-infrared band
$10$--$10^3$ sec after the GWs for an sGRB~170817A-like event,
which is brighter than the early macronova (or kilonova) emission 
and detectable by LSST in the near future.
We also propose a new method to constrain the unknown jet properties, such as
the Lorentz factor, opening angle, emission radii, and jet launch time,
by observing both the forward and counter jets.
To scrutinize the counter jets,
space GW detectors like DECIGO are powerful in forecasting
the merger time ($\lesssim 1$ sec) and position ($\lesssim 1$ arcmin) 
($\sim$ a week) before the merger.
\end{abstract}
\subjectindex{E32, E01, E02, E37}

\maketitle


\section{Introduction}

A short Gamma-Ray Burst (sGRB) is believed to arise from a relativistic jet.
However the physical properties of the sGRB jets are still enigmatic
\cite[e.g.,][]{Zhang_Meszaros04,Nakar07}.
For example, the jet opening angles are not determined for most sGRBs.
As far as we know, there are only five sGRBs for which the jet opening angle is determined 
by a break in the afterglow light curve.
They are GRB~140903A \cite{Troja16}, 130603B \cite{Fong14}, 
111020A \cite{Fong12}, 090426 \cite{Nicuesa11}, and 051221 \cite{Soderberg06}.
The derived jet opening angles $\Delta \theta$ for these sGRBs are 
$\sim 5^\circ$, $4^\circ$--$8^\circ$, $3^\circ$--$8^\circ$, $4.4^\circ$, and $5.7^\circ$--$7.3^\circ$, respectively,
with the mean value of $\sim 6^\circ$. 
On the other hand, there exist seeven sGRBs for which a jet break was not detected during the afterglow observation and hence only the lower limits on the jet opening angles are obtained,
ranging from $4^\circ$ to $25^\circ$ \cite{Fong15}.
Using these data, \citet{Fong15} obtained a mean value $\Delta\theta=16^\circ\pm10^\circ$.
The other properties such as the exact value of the bulk Lorentz factor,
emission radius/mechanism, composition, and structure
are also still unclear.

One critical reason for our ignorance of sGRB jets is that
the counter jet of an sGRB has not been detected so far.
For active galactic nuclei (AGNs),
the counter jet is sometimes observed, 
leading to constraints on its Lorentz factor and viewing angle.
\cite[e.g.,][]{Begelman+84}.
Similar arguments are also applied to Galactic microquasars \cite[e.g.,][]{Mirabel_Rodriguez98}.
For long GRBs, the off-axis jet emission was first considered in the context of
the diversity of GRB appearance
and a possible origin of X-ray flashes 
\cite[e.g.,][]{Ioka2001,Yamazaki2002,Yamazaki2003c,Yamazaki2003b,Yamazaki2004}.
\citet{Yamazaki2003} claimed that the counter-jet emission can be observed as
``delayed flashes'' in the UV-optical bands,
although detection has not succeeded yet
because relativistic debeaming makes the emission very faint.
Since there is no confirmed off-axis sGRB,
90--99\% of sGRBs have not been observed so far.

The recent discoveries of a double neutron star (NS) merger GW170817 
by gravitational waves (GWs) \cite{LVC170817a}
and the associated sGRB~170817A \cite{GW-GRB17,Fermi/GBM17,Integral17}
have made a breakthrough on the issue.
The distance to sGRB~170817A is very close at $\sim$ 40~Mpc \cite{Abbot2017}
while the nearest sGRB so far is around ten times further away.
The viewing angle ($\theta_v$) of the sGRB~170817A jet, which is the angle between the line of sight and the orbital axis of the binary,
is constrained by the GWs as $\theta_v \lesssim 32^\circ$ \cite{LVC170817a,GW170817-H0}.
More accurate determination of $\theta_v$ is expected with an increase of the 
sensitivity and the number of GW detectors such as KAGRA \cite{Abbott+16}.
The observed weakness of the prompt sGRB can be interpreted by
an off-axis emission from a canonical sGRB jet \cite{GW-GRB17,Ioka2017,Granot+17,Wang+17},
although there remain other possibilities such as
a structured jet
\cite[e.g.,][]{Lamb_Kobayashi17,Jin+17,Lazzati+17,Kathirgamaraju+17},
a breakout emission from a mildly-relativistic cocoon
\cite[e.g.,][]{Nagakura+15,Kasliwal+17,Gottlieb+17,Bromberg+17},
a low-luminosity sGRB population
\cite[e.g.,][]{Murguia-Berthier+17,Zhang+17,Yue+17},
a giant flare
\cite{Salafia+17,Tong17},
a prompt jet emission scattered by the cocoon
\cite{Kisaka+15,Kisaka+17},
and a spiral electron jet \cite{Fargion+17}.
The late-time X-ray and radio data can be also explained by
the off-axis afterglow emission from a canonical sGRB jet
\cite{Alexander+17,Swift-NuSTAR17,Hallinan+17,Margutti+17,Troja+17}.

In this paper we apply and extend \citet{Yamazaki2003} to sGRB~170817A-like events.  
The emission of the prompt counter jet is delayed from that of the forward jet by $10$--$10^3$ sec
and observable as an optical-infrared burst.
We show that it is brighter than the macronova (or kilonova) emission
from the $r$-process radioactive matter ejected by the NS merger
\cite{Coulter+17,Tanaka+17b,Utsumi+17,Tominaga+17,Drout+17,Swift-NuSTAR17,Arcavi+17,Smartt+17,
Shappee+17,Pian+17,Kasen+17,Kasliwal+17,Tanvir+17,Kilpatrick+17,Soares-Santos+17,Cowperthwaite+17,
Nicholl+17,Chornock+17,Valenti+17,Diaz+17,McCully+17,Buckley+17}.
Then we will get four times, that is, the start and end times of the forward jet emission and the counter jet emission, 
which enable us to get the Lorentz factor of the jet $\gamma$, 
the opening angle $\Delta \theta$, and 
the starting and ending radii of the jet emission
as functions of the jet launch time $\tau_j$.    
For a viewing angle of the jet $\theta_v\sim \pi/2$, 
we can determine $\tau_j$.
Once we obtain the probability distribution of $\tau_j$,
we can determine the distributions of all the parameters (see Appendices~B and C).  
These observed parameters constrain various emission mechanisms of sGRBs.

This paper is organized as follows. 
In \S~2, a model of the forward and the counter jet is given.
In \S~3, we compare the observed flux of the prompt counter-jet emission 
with that of a macronova.
Section~4 is devoted to discussions.
The details of the calculations are given in Appendices A, B, and C.

%

\section{Model of prompt emission from a counter jet}

Emission from a counter jet is less beamed than that from the forward jet.
Assuming that the counter jet has the same physical quantities 
as the forward jet except for the direction,
the observed peak frequency of the counter-jet emission is typically
ultraviolet or soft X-rays,
\begin{eqnarray}
\nu_{\rm peak}^{\rm (c)} \sim \frac{1}{4 \gamma^2} \nu_{\rm peak}^{\rm (f)} 
\sim 30\,{\rm eV} 
\left(\frac{\gamma}{10^2}\right)^{-2}
\left(\frac{\nu_{\rm peak}^{\rm (f)}}{1\,{\rm MeV}}\right),
\end{eqnarray}
where $\nu_{\rm peak}^{\rm (f)}$ is a typical peak frequency of the 
on-axis forward-jet emission, and $\gamma$ is the Lorentz factor of the jets.
The counter-jet emission is rather insensitive to the viewing angle.
The emission is delayed after the forward-jet emission by
\begin{eqnarray}
T_{\rm start/end}^{\rm (c)} \sim \frac{2 r_0}{c} \sim 2\times 10^2\,{\rm s} 
\left(\frac{r_0}{3\times 10^{12}\,{\rm cm}}\right),
\label{eq:timescale}
\end{eqnarray}
where the emission radius $r_0$ depends on the emission mechanism
and is 
$\sim \gamma^2 c \Delta t \sim 3 \times 10^{12}\,{\rm cm}\,(\gamma/10^2)^2
(\Delta t/10\,{\rm ms})$ for the case of internal shocks
with variability timescale $\Delta t$.\footnote{
The light curve of the counter-jet emission is rather smooth.
Some sGRBs, which are the on-axis emission of the forward jet,
consist of multiple pulses whose separation is on the order of $\Delta t\sim r_0/c\gamma^2$.
Even in this case, each pulse of the corresponding counter-jet emission has a much longer width
$\Delta T^{\rm (c)}=T_{\rm end}^{\rm (c)}-T_{\rm start}^{\rm (c)}\sim r_0/c$, so that
all the counter-jet pulses overlap with each other, resulting in a smooth light curve with a single peak.
}

We consider a simple model for the emission from a relativistic jet \cite{Yamazaki2003}
(see also \cite{Lazzati+17b}).
The jet radiates photons at radii from $r_0$ to $r_{\rm e}$. We assume an optically thin, 
instantaneous thin-shell emission, i.e., the cooling timescale is much shorter than the other timescales.
We introduce a spherical coordinate system $(t,r,\theta,\phi)$
in the Lab frame, where the $\theta=0$ axis points 
toward the detector at $r=D$, the central engine is located at $r=0$,
and $t=0$ is the end time of the GW emission.
The viewing angle $\theta_v$ of the forward jet is defined by the angle between 
the forward-jet axis and the line of sight while that of the counter jet is $\theta_v+\pi$.  
(see Figure~1).
Both jets are launched from $r=0$ at $t=\tau_j$, 
where jets launched before $t=\tau_j$ collide with the merger ejecta
and are engulfed in the cocoon \cite{Nagakura+15}.\footnote{
The starting time of the jet launch is earlier than $t=\tau_j$
because it takes time for a jet to penetrate the merger ejecta
\citep[e.g.,][]{Nagakura+15,Ioka2017}.}
The emitting shock front moves radially from $r=r_0$  at $t=\tau_j+t_0$  with 
the Lorentz factor $\gamma=1/\sqrt{1-\beta^2}$.

The observer time $T=0$ was chosen when the GW detection ends (see Table~1).
Then, for a single pulse, we get the observed flux per unit frequency 
at the observer time $T$ as\footnote{
In this paper, we neglect
the effect of cosmological expansion for simplicity, 
because the source we consider is located much closer than 1~Gpc (corresponding to $z\approx0.2$). }
\beqa
F_{\nu}(T)={{2 {r_0}^{2}}\over{\beta D^2 (r_0/c\beta)}}
\int dt A(t) 
{{[\gamma(1-\beta \cos\theta(T))]}\over
{[\gamma(1-\beta \cos\theta(t))]}}
{{\Delta \phi(t) f[\nu \gamma(1-\beta \cos \theta(t))]}\over
{[\gamma(1-\beta \cos\theta(t))]^2}},
\label{eq:jetthin}
\eeqa
where
$f(\nu')$ represents the spectral shape,
\begin{eqnarray}
1-\beta\cos\theta(T)&=&\frac{c\beta}{r_0}(T-T_0)~~, 
\label{eq:thetaT1} \\
1-\beta\cos\theta(t)&=&\frac{T-T_0}{t-T_0}~~,
\label{eq:thetaT2}
\end{eqnarray}
and $T_0=\tau_j+t_0-r_0/c\beta$.
The derivation of Eq.~(\ref{eq:jetthin}) is given in Appendix~A.
For the counter-jet emission, 
$\theta(t)$ varies from $\pi+\theta_v+\Delta \theta$
to $\pi+{\rm max}\{0,\theta_v-\Delta\theta\}$,
and the polar (half-)angle of the emitting region
$\Delta \phi(t)$ is given as
\begin{equation}
\Delta\phi(t)=
\left\{
\begin{array}
{c@{} l@{}}
\pi & (\theta_v<\Delta\theta \>\>{\rm and}\>\> 
\pi<\theta(t)\leq \pi+\Delta\theta-\theta_v )\\
\cos^{-1}\left[
\f{\cos \Delta \theta + \cos \theta(t) \cos \theta_v}
{-\sin \theta_v \sin \theta(t)}\right]
& ({\rm otherwise}) 
\end{array} \right. .
\label{eq:delta_phi}
\end{equation}
The normalization of the emissivity $A(t)$ 
is determined by the hydrodynamics and microphysics.
Here we set the following functional form,
\beqa
A(t)=A_0\left(\frac{t-T_0}{r_0/c\beta}\right)^{-2}
H(t-\tau_j-t_0) H(\tau_j+t_e-t),
\label{eq:At}
\eeqa
where $H(x)$ is the Heaviside step function 
for the emission to start at $t=\tau_j+t_0$ and end at $t=\tau_j+t_e$,
and the time dependence $(t-T_0)^{-2}$ yields
constant energy release at each distance $r$.
We define $\kappa \equiv t_e/t_0$, which is larger than unity.
The precise form of $A(t)$ does not change our conclusion so much.
The starting time and the ending time of the prompt counter-jet emission are given as
\begin{eqnarray}
T_{\rm start}^{\rm (c)}&=&T_0+({r_0}/{c\beta})
\left[1+\beta\cos(\theta_v+\Delta\theta)\right],
\label{eq:tstart:uvf}
\\
T_{\rm end}^{\rm (c)}&=&T_0+[({r_0}/{c\beta})+t_e-t_0]
\left[1+\beta\cos(\max[0,\theta_v-\Delta\theta])\right],
\label{eq:tend:uvf}
\end{eqnarray}
respectively.
The spectrum of GRBs is well approximated by 
the Band function \cite{Band1993}. 
In order to have a spectral shape similar to the Band function,
we assume the following form of the spectrum in the comoving frame:
\beqa
f(\nu')=(\nu'/\nu'_0)^{1+\alpha_B}
[1+(\nu'/\nu'_0)^s]^{(\beta_B-\alpha_B)/s},
\label{eq:spectrum}
\eeqa
where $\alpha_B$ ($\beta_B$) is the low- (high-)energy power-law index,
and $s$ describes the smoothness of the transition.
Typical values are $\alpha_B\sim -1$ and $\beta_B\sim -3$ for GRBs \cite{Preece2000}.
Now equations (\ref{eq:jetthin}), (\ref{eq:At}), and (\ref{eq:spectrum})
are the basic equations to calculate the flux of a single pulse,
which depends on the following parameters:
$\gamma \gg 1$, $\theta_v$, $\Delta \theta$, $r_0$, 
$\gamma \nu_0'$,  $\alpha_B$, $\beta_B$, $s$,
$D$, $A_0$, $t_0$, $\tau_j$, and $\kappa$.

\begin{figure}[t]
\vspace{1 cm}
\begin{center}
\includegraphics[width=55mm, angle=270]{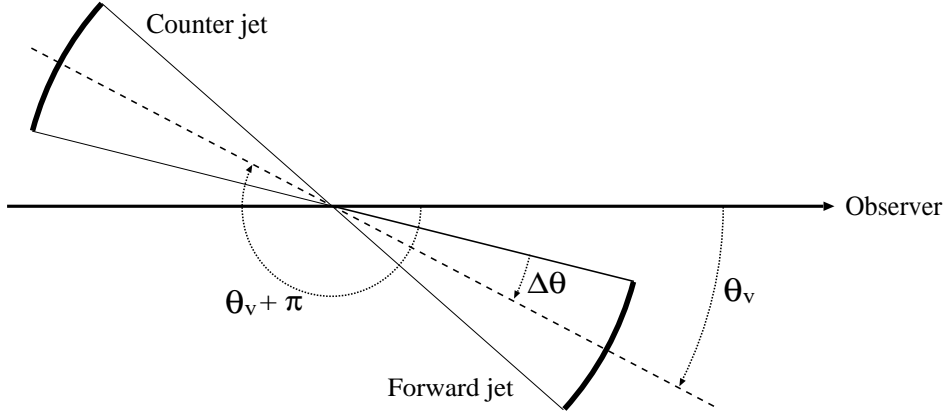} 
\caption{
Geometry of the forward jet, the counter jet and the observer.    }
  \end{center}
 \label{fig1}
\end{figure}

\begin{table}[t]
\begin{center}
\begin{tabular}{ ll | ll  }
\hline
\multicolumn{2}{c|}{Time coordinate in Lab frame}  
& \multicolumn{2}{c}{Observer time}  \\ 
\hline \hline
$t=0$ & GW emission ends. & $T=0$ & GW detection ends. \\
$t=\tau_j$ & Relativistc jets launched from $r=0$.
& & \\
$t=\tau_j+t_0$ & Jet emissions start.
& $T=T^{\rm (f)}_{\rm start}$ & Forward-jet emission starts.\\
 & & $T=T^{\rm (c)}_{\rm start}$ & Counter-jet emission starts.\\
$t=\tau_j+t_e$ & Jet emissions end.
& $T=T^{\rm (f)}_{\rm end}$ & Forward-jet emission ends.\\
 & & $T=T^{\rm (c)}_{\rm end}$ & Counter-jet emission ends.\\
 \hline
\end{tabular}
\end{center}
\caption{Time coordinate, $t$, in Lab frame and the observer time $T$. }
\label{table1}
\end{table}

Hereafter we choose the following fiducial parameters:
$\gamma=100$, $\Delta\theta=20^\circ$, 
$h\gamma\nu'_0=500\,{\rm keV}$,
$r_0=1\times10^{12}\,{\rm cm}$, $\alpha_B=-1$, $\beta_B=-3$, $s=1$,
and $t_0=r_0/c\beta$.
We adopt $\kappa=1.3$
since most GRB pulses (forward-jet emissions) rise more quickly than they decay 
\cite{Norris1996}.
We determine the amplitude $A_0$ so that
the isotropic-equivalent gamma-ray energy of the forward-jet emission 
$E_{\rm iso,on}$ satisfies
$(1/2)(\Delta\theta)^2E_{\rm iso,on}=5\times10^{50}{\rm erg}$
when it is viewed on-axis ($\theta_v=0$).
For our fiducial parameters, we obtain  $E_{\rm iso,on}=8.2\times10^{51}$erg.

Figure~2 shows {\it r}-band ($\nu_r=4.6\times10^{14}$Hz) light curves 
of the counter-jet emission
for the fiducial parameters 
by varying $\theta_v$ ($=0^\circ$, $20^\circ$, $30^\circ$, and $40^\circ$). 
In cases of $\theta_v<\Delta\theta$, each light curve shows 
a constant flux phase because we use a top-hat jet with uniform surface brightness.
Furthermore, the peak flux does not depend greatly on $\theta_v$  \cite{Yamazaki2003}.
This is because
the value of $\theta(t)$ ranges between 
$\pi+{\rm max}\{0,\theta_v-\Delta\theta\}$ and
$\pi+\theta_v+\Delta\theta$,
and $(c\beta/r_0)(t-T_0)\sim 1$,
so that
the beaming factors are
$1/\gamma (1-\beta\cos\theta(t))
\sim 1/\gamma (1-\beta\cos\theta(T))
\sim 1/2\gamma$
in Eq.~(\ref{eq:jetthin}).
For the fiducial parameters, the counter-jet emission starts at $\sim 1$~min after the merger event
with a peak flux of about 23~mag if $D=40$~Mpc.
As shown in the figure, the prompt counter-jet emission starts slightly earlier for larger $\theta_v$.
Assuming $\beta\approx1$, $\theta_v<1$ and $\gamma^{-2} \ll \Delta \theta<1$, we obtain for $\theta_v>\Delta\theta$,
\begin{eqnarray}
T_{\rm start}^{\rm (c)}-\tau_j&=&\frac{2r_0}{c}
\left[1-\frac{(\theta_v+\Delta\theta)^2}{4}\right],
\label{eq:tstart2}\\
T_{\rm end}^{\rm (c)}-\tau_j&=&\frac{2\kappa r_0}{c}
\left[1-\frac{(\theta_v-\Delta\theta)^2}{4}\right],
\label{eq:tend2}
\end{eqnarray}
where we use the approximation\footnote{
The approximation of $\cos\theta\sim 1-\theta^2/2$ is
good for $\theta <1$. For example, for $\theta=1~{\rm rad}=57.3^\circ$, 
the approximation gives $\cos(\theta)=0.5$ while $\cos(60^\circ)=0.5$.
}
$\cos\theta\sim 1-(\theta^2/2)$ for $\theta <1$.
We note here that $(T_{\rm end}^{\rm (c)}-T_{\rm start}^{\rm (c)})/(T_{\rm start}^{\rm (c)}-\tau_j)\sim
\kappa-1 = 0.3$,
neglecting the second terms of the above equations that include $\theta_v$ and $\Delta\theta$.

In Fig.~\ref{fig3}, one can see the parameter dependence of the counter-jet emission.
From Eq.~(\ref{eq:jetthin}),
 the peak flux can be estimated as
$F_{\nu}\approx (2r_0^2 \gamma^2/\beta D^2) A_0 \Delta \phi f/(2\gamma^2)^2$
where $\Delta\phi$ is approximately of order unity, and 
$f\propto \nu^{1+\alpha_B}$ because we consider the case with $\nu\ll\gamma\nu'_0$ \cite{Yamazaki2003}.
Since $A_0\propto E_{\rm iso,on}/r_0^3$ in the case of Eq.~(\ref{eq:At}), 
we find
$F_\nu\propto(E_{\rm iso,on}/r_0\gamma^2)\nu^{1+\alpha_B}$.
Therefore, the counter-jet emission becomes brighter for smaller $\gamma$ and $r_0$.
As long as $\alpha_B\approx-1$, the flux is insensitive to observation bands in the UV, optical,
and infrared ranges.
   
   As shown in Appendix~B, from the observed four time values, that is,  
$T_{\rm start}^{\rm (c)}$, $T_{\rm end}^{\rm (c)}$, $T_{\rm start}^{\rm (f)}$, and $T_{\rm end}^{\rm (f)}$,    
we can determine the Lorentz factor of the jet $\gamma$, the opening angle $\Delta \theta$, and
the starting and ending emission radii of the jet from the central engine     
as functions of the jet launch time $\tau_j$ 
by using the viewing angle $\theta_v$ which is determined by the GWs.
For a jet viewing angle of  $\theta_v\sim \pi/2$, 
the jet launch time $\tau_j$ can be determined as shown in Appendix~C. 
Once the probability distribution of the jet launch time $\tau_j$ is obtained,
we can in principle estimate the distributions of
the Lorentz factor $\gamma$, 
the opening angle $\Delta\theta$, 
the starting radius $r_0$, and the ending radius $r_e$ of the prompt emissions. 
These observed parameters constrain various emission mechanisms of sGRBs.
\begin{figure}[t]
\begin{center}
\includegraphics[width=90mm]{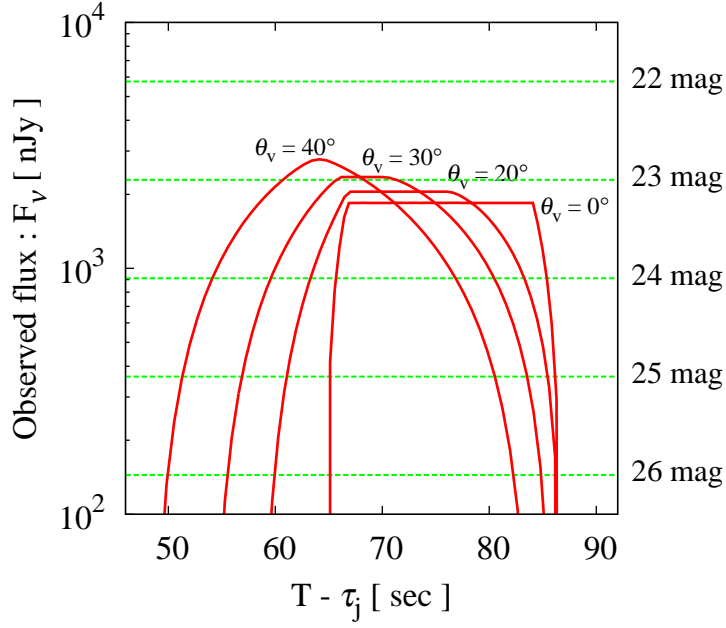}
\vspace{-2.0 cm}
\caption{
Light curves of the counter-jet emission in the {\it r}-band for the fiducial parameters
($\gamma=100$, $\Delta\theta=20^\circ$, 
$h\gamma\nu'_0=500\,{\rm keV}$,
$r_0=1\times10^{12}\,{\rm cm}$, $\alpha_B=-1$, $\beta_B=-3$, $s=1$,
$t_0=r_0/c\beta$,  $\kappa=1.3$,
and $E_{\rm iso,on}=8.2\times10^{51}$erg)
by varying $\theta_v$ ($=0^\circ$, $20^\circ$, $30^\circ$, and $40^\circ$, from right to left). 
The source is located at $D=40$~Mpc.
The observer time $T=0$ is set when the GW detection ends, and
$\tau_j$ is the jet launch time (see text for details).
    }
  \end{center}
 \label{fig2}
\end{figure}

\section{Early-epoch macronova emission}

Since an sGRB is likely to be associated with a binary NS merger,
the mass ejection at the merger leads to macronova emission via radioactivity,
which may hide the counter-jet emission.
In this section, following Kisaka~et~al. \cite{Kisaka2015},
we calculate the flux of a very early macronova to compare it with the counter-jet emission.

We start with the radioactive heating rate $\epsilon_{\rm th}{\dot \varepsilon}_r$, where
$\epsilon_{\rm th}$ is the thermalization factor, describing the fraction of the decay energy deposited to the ejecta.
We adopt 
${\dot \varepsilon}_r(t) = 2\times 10^{10} (t/1\,{\rm day})^{-1.3}$~erg~s$^{-1}$g$^{-1}$,
which gives a reasonable agreement 
with nucleosynthesis calculations for a wide range of 
the electron fraction \cite{Korobkin2012,Rosswog2014,Wanajo2014},
and we assume $\epsilon_{\rm th} =1$ since we consider the very early epoch
\citep{Barnes+16,Hotokezaka+16}.
The dynamical ejecta are assumed to have 
a homologously expanding density profile
$\rho(t,v)\propto t^{-3}v^{-3.5}$ with $v_{\rm min}<v<v_{\rm max}$ \cite{Kisaka2015}
and a total mass $M_{\rm ej}$.
Then, the blackbody temperature in the thin-diffusion regime, which is applicable to the very early phase, is (see Eqs.~(18) and (A.16) of \cite{Kisaka2015}):
\begin{eqnarray}
T_{\rm BB} (t)
&\approx& \left(\frac{\epsilon_{\rm th}{\dot \varepsilon}_r(t) t \rho(t,v_{\rm max})}{a}\right)^{1/4} \nonumber\\
&=& 1.2 \times10^6~{\rm K}
\left(\frac{M_{\rm ej}}{0.03M_{\odot}}\right)^{1/4}
\left(\frac{v_{\rm min}}{0.1c}\right)^{0.125}
\left(\frac{v_{\rm max}}{0.4c}\right)^{-0.875}
\left(\frac{t}{10^2~{\rm s}}\right)^{-0.825},
\end{eqnarray}
where $a=7.56\times10^{-15}~{\rm erg}~{\rm cm}^{-3}~{\rm deg}^{-4}$ is the radiation density constant.
The bolometric luminosity is given by 
(see Eqs.~(6), (17), (19) and (A.17) of \cite{Kisaka2015})
\begin{eqnarray}
L_{\rm bol}(t)&\approx&
4\pi(v_{\rm max}t)^2
\left(\frac{ct}{\kappa \rho(t,v_{\rm max})}\right)^{1/2}
\rho(t,v_{\rm max})\epsilon_{\rm th}{\dot \varepsilon}_r(t) \nonumber\\
&=&1.5 \times10^{42}~{\rm erg}~{\rm s}^{-1}
\left(\frac{\kappa}{10~{\rm cm}^2~{\rm g}^{-1}}\right)^{-1/2}
\left(\frac{M_{\rm ej}}{0.03M_{\odot}}\right)^{1/2} \nonumber\\
&& \ \ \ \times
\left(\frac{v_{\rm min}}{0.1c}\right)^{0.25}
\left(\frac{v_{\rm max}}{0.4c}\right)^{0.25}
\left(\frac{t}{10^2~{\rm s}}\right)^{-0.3}~~,
\end{eqnarray}
where $\kappa$ is the opacity of the ejecta.
Hence, the observed flux of the early-phase macronova emission is calculated as
\begin{eqnarray}
F_\nu (t) &=&\frac{L_{\rm bol}(t)}{4\pi D^2}
\frac{15}{\pi^4\nu}\left(\frac{h\nu}{kT_{\rm BB}}\right)^3
g(h\nu/kT_{\rm BB})
\nonumber\\
&=&4.6~{\rm nJy} 
\left(\frac{\kappa}{10~{\rm cm}^2~{\rm g}^{-1}}\right)^{-1/2}
\left(\frac{M_{\rm ej}}{0.03M_{\odot}}\right)^{-1/4} 
\left(\frac{v_{\rm min}}{0.1c}\right)^{-0.125}
\left(\frac{v_{\rm max}}{0.4c}\right)^{2.875} \nonumber\\
&& \ \ \ \ \times
\left(\frac{\nu}{\nu_r}\right)^2
\left(\frac{D}{40~{\rm Mpc}}\right)^{-2}
\left(\frac{t}{10^2~{\rm s}}\right)^{2.175}
g(h\nu/kT_{\rm BB})~~,
\label{eq:flux_macronova}
\end{eqnarray}
where $g(y)=y(e^y-1)^{-1}$ and $\nu_r=4.6\times10^{14}$Hz is the {\it r}-band frequency.
One can see that $g(h\nu_r/kT_{\rm BB})\approx1$ 
(i.e., the Rayleigh-Jeans limit) for $t<10^3$~s. 
In Figure~3, we show macronova light curves for two cases:
$\kappa=0.3$~cm$^{2}$~g$^{-1}$ and $M_{\rm ej}=0.02~M_\odot$ (denoted as ``blue macronova''), and
$\kappa=10$~cm$^{2}$~g$^{-1}$ and $M_{\rm ej}=0.03~M_\odot$ (denoted as ``red macronova'')
\cite{Coulter+17,Tanaka+17b,Utsumi+17,Tominaga+17,Drout+17,Swift-NuSTAR17,Arcavi+17,Smartt+17,
Shappee+17,Pian+17,Kasen+17,Kasliwal+17,Tanvir+17,Kilpatrick+17,Soares-Santos+17,Cowperthwaite+17,
Nicholl+17,Chornock+17,Valenti+17,Diaz+17,McCully+17,Buckley+17}.
In both cases, we set $v_{\rm min}=0.1c$, $v_{\rm max}=0.4c$, and $D=40$~Mpc.
In drawing the thin green lines in Figure~3, we assume $t\approx T$ since the merger ejecta expands
nearly isotropically with sub-relativistic speeds.
The jet launch time $\tau_j$ is smaller than $\sim 2$~sec, which is inferred
 from the observation of GW170817/sGRB~170817A. This event showed that the gamma-ray emission
started $\sim 1.7$~sec after the end of the GW emission \cite{Ioka2017}.
As long as $\tau_j\ll10^2$~s, the choice of $\tau_j$ does not affect the macronova light curves 
for $T>10^2$~s, which we are interested in.
Finally, we note that the dependencies of the macronova flux on $M_{\rm ej}$ and $v_{\rm min}$
are very weak.

\begin{figure}[t]
\begin{center}
\includegraphics[width=170mm]{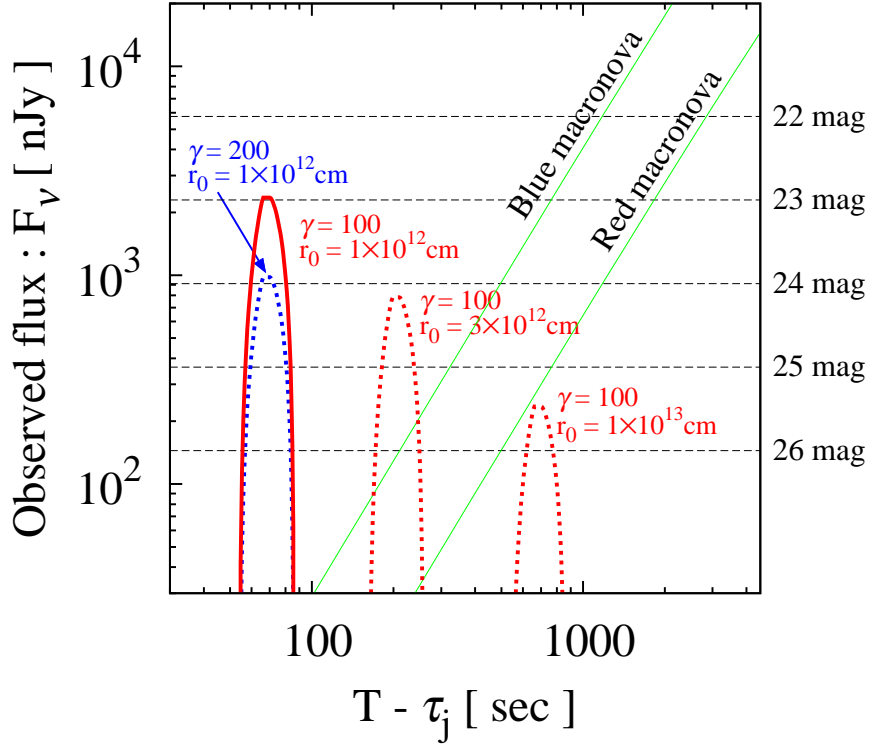} 
\vspace{-0.5 cm}
\caption{
Light curves of the counter-jet emission in the {\it r}-band. 
The source is located at $D=40$~Mpc.
The thick red line is  for the fiducial parameters
($\gamma=100$, $\Delta\theta=20^\circ$, 
$h\gamma\nu'_0=500\,{\rm keV}$,
$r_0=1\times10^{12}\,{\rm cm}$, $\alpha_B=-1$, $\beta_B=-3$, $s=1$,
$t_0=r_0/c\beta$,  $\kappa=1.3$,
and $E_{\rm iso,on}=8.2\times10^{51}$erg)
with $\theta_v=30^\circ$.
The observer time $T=0$ is set to be the ending time of the GW detection, and
 $\tau_j$ is the jet launch time (see text for details).
Other cases in which one of the fiducial parameters is changed are also shown (dashed lines).
Two thin green lines represent the macronova emission from the merger ejecta with
the opacity $\kappa=0.3$~cm$^{2}$~g$^{-1}$ and the total mass $M_{\rm ej}=0.02 M_\odot$ (blue macronova)
and $\kappa=10$~cm$^{2}$~g$^{-1}$ and $M_{\rm ej}=0.03 M_\odot$ (red macronova).
In both cases, we adopt
the minimum velocity $v_{\rm min}=0.1c$ and the maximum velocity $v_{\rm max}=0.4c$.
Note that the macronova emission in the very early epoch scales as
$F_{\nu}\propto \kappa^{-1/2}M_{\rm ej}^{-1/4}v_{\rm min}^{-0.125}v_{\rm max}^{2.875}T^{2.175}\nu^2$.
    }
  \end{center}
 \label{fig3}
\end{figure}


\section{Discussions}

We have calculated the {\it r}-band light curves of the counter-jet emission and the 
early-phase macronova associated with sGRBs.
We have shown that the counter-jet emission is observable 
about $10$--$10^3$~s after the binary NS merger 
if the emission radius is comparable to the internal shock radius $r_0 \sim 10^{11}$--$10^{13}$~cm
and the bulk Lorentz factor is not too large $\gamma \lesssim 300$,
because the emission becomes brighter than the macronova.
For our fiducial parameters, 
$r_0=1\times 10^{12}$--$3\times10^{12}$~cm and $\gamma=50$--200, the counter jet emission
has an apparent magnitude brighter than $\sim24$~mag if the source is located within 40~Mpc.
As shown in \S~2, the observed flux of the counter-jet emission only weakly depends on the observation band if the low-energy spectral index 
is typical ($\alpha_B\approx-1$).
On the other hand, the early macronova emission in the optically thick regime is brighter for higher frequency bands
[see eq.~(\ref{eq:flux_macronova})].
For example, the {\it g}-band flux is about 0.7~mag brighter than the {\it r}-band flux.
Hence, longer-wavelength observations are favorable for the detection of the counter-jet emission.

In order to detect the counter-jet emission,
rapid follow-up observations with short integration time are required
because the relevant timescale is about 1~min 
[see Eqs.~(\ref{eq:timescale}), (\ref{eq:tstart2}) and (\ref{eq:tend2})].
In the near future, the Large Synoptic Survey Telescope (LSST)\footnote{
https://www.lsst.org}
will have sensitivity of 24.7 and 25.0~mag for 30~sec
exposure in the {\it r}- and {\it g}-bands, respectively \cite{LSST2011},
so that the counter-jet emission would be detectable within $\sim 90$~Mpc
for our fiducial model parameters.

When DECIGO (Decihertz Interferometer Gravitational Wave
Observatory) \cite{Seto2001} is launched in 2030s,
we can precisely determine the source direction and the merger time
long before the NS coalescence \cite{Takahashi2003}. 
The main reason is the accuracy of the angular resolution. An NS-NS merger emits
$\sim$ 0.1~Hz GWs $\sim1$~year before the final merger. 
The orbit of the DECIGO is near the Earth orbit around the Sun. 
Since the wavelength of a $\sim0.1$~Hz gravitational wave is $\sim10^{11}$~cm
and the orbital radius of DECIGO is $\sim1$~au, 
the angular resolution $\delta\theta$ is given by
\begin{equation}
\delta\theta\sim \frac{10^{11}~{\rm cm}}{10^{13}~{\rm cm}}\frac{1}{SNR}\sim 3~{\rm arc min}\frac{10}{SNR}~.
\end{equation}
According to \cite{Takahashi2003}, the sky position and the merger time can be predicted
with accuracies $\sim1$~arcmin and $\sim 1$~sec, respectively, roughly a week before the merger.
Then we can prepare the electromagnetic telescopes and satellites for the NS-NS merger events,
similarly to a solar eclipse.
Therefore DECIGO will make it much easier to detect the prompt counter-jet emission.

We can also calculate the X-ray flux of the counter-jet emission.
For fiducial parameter sets, we obtain 
$\nu F_\nu\sim5\times10^{-15}(r_0/10^{12}{\rm cm})^{-1}(D/40~{\rm Mpc})^{-2}$erg~s$^{-1}$cm$^{-2}$
at $h\nu=1$~keV.
If we adopt $h\gamma\nu'_0=4\,{\rm MeV}$ and $\beta_B=-2.5$ with the other parameters being fiducial,
the X-ray flux becomes 20 times larger, but it is still below the sensitivity of {\it Swift}/XRT 
with a short exposure of $\sim10^2$~s ($\sim4\times10^{-13}$erg~s$^{-1}$cm$^{-2}$).

There are several other emission components which potentially dominate
over the counter-jet emission.
One is the afterglow from the forward jet \cite{Lamb_Kobayashi17,Xiao+17,Granot+17b}. 
However, in our case, the off-axis afterglow peaks at a later time,
so that the emission is very weak in the very early epoch ($10^{2-3}$~s).
The early afterglows caused by macronova ejecta are also dim \cite{Asano2017}.
Another is the optical counterpart of the temporally extended emission
associated with sGRBs \citep[e.g.,][]{Kisaka+15}.
The optical flux is weak enough if the emission radii are smaller than $\sim10^{11}$~cm
due to self-absorption or absorption by the merger ejecta.
It is also possible that the ``neutron precursor'', which is powered by the $\beta$-decay
of free neutrons in the outermost macronova ejecta, becomes brighter than the counter-jet emission
\cite{Metzger+15}.
When these additional components dominate over the counter-jet emission, the radius of the
extended emission can be constrained,
which may become another interesting topic.

The counter jet also gives rise to a radio afterglow 
when it is decelerated by the surrounding medium.
It appears as a late-time bump in the light curve, exceeding the forward-jet afterglow
\cite{Granot2003,Wang2009,Zhang2009},
when the counter jet slows down to be non-relativistic.
The time for the bump in the light curve is estimated as
$T_{\rm bump}\approx9\times10^3(E_{\rm K,iso}/10^{52}{\rm erg})^{1/3}
(n_0/10^{-3}{\rm cm}^{-3})^{-1/3}{\rm days}$, where
$E_{\rm K,iso}$ and $n_0$ are the isotropic-equivalent kinetic energy of the counter jet and
the ambient matter density, respectively \cite{Zhang2009}.
However, the initially non-relativistic merger ejecta, which was considered in \S~3, is also
associated with the binary NS merger events.
It produces bright synchrotron emission at radio wavelengths
once it is decelerated \cite{Nakar2011,Piran2013,Takami2014}.
For example, if the dynamical ejecta has $M_{\rm ej}=0.02~M_\odot$ and $v_{\rm ej}=0.3c$,
the radio emission peaks at $\sim6\times10^3$~days for $n_0\sim10^{-3}$cm$^{-3}$
\cite{Alexander+17}, and its flux is much larger than that of the counter-jet afterglow
(see Extended Data Figure~2 of \cite{Troja+17}).

The starting time $T_{\rm start}^{\rm (c)}$ and the ending time $T_{\rm end}^{\rm (c)}$ of the 
counter-jet emission are roughly estimated
as $T_{\rm start}^{\rm (c)}\sim T_{\rm end}^{\rm (c)}\sim (2 r_0/c) +\tau_j$. 
Therefore,  observations of the counter-jet emission directly give us 
an order-of-magnitude estimate of the emission radius $2 r_0+c\tau_j$.
In addition, if the prompt emission of the forward jet is detected,
the starting and ending times of the forward-jet emission,
$T_{\rm start}^{\rm (f)}$ and $T_{\rm end}^{\rm (f)}$, further constrain the model parameters.
%
Remembering that
the viewing angle $\theta_v$ will be determined by the analysis of the GW
from the binary NS merger,
 the four observables, $T_{\rm start}^{\rm (c)}$, $T_{\rm end}^{\rm (c)}$,
$T_{\rm start}^{\rm (f)}$ and $T_{\rm end}^{\rm (f)}$, are functions of five unknown
model parameters, $\gamma$, $\Delta\theta$, $r_0$, $\kappa$ and $\tau_j$.
Therefore, in principle, we can determine the four model parameters as a function of $\tau_j$.
If the viewing angle of the jet is $\theta_v\sim \pi/2$, we can determine $\tau_j$.
Once we obtain the probability distribution of $\tau_j$,
we can estimate the distributions of
the Lorentz factor $\gamma$,
the opening angle $\Delta\theta$, 
the starting radius $r_0$, and the ending radius $r_e$ of the emissions 
(see Appendix~C for details).
These observed parameters constrain various models of prompt sGRBs
and even fundamental physics
\cite{Wang+17,Shoemaker17,Sakstein17,Ezquiaga17,Creminelli17}.

The prompt emission mechanism could be different from  internal shocks.
If the prompt emission arises from the photosphere \cite{Meszaros_Rees00,Kisaka+17},
the emission radius is small ($\sim 10^{9}$--$10^{10}$~cm)
and may be surrounded by or close to the merger ejecta.
In this case, the counter-jet emission may be blocked by the optically thick merger ejecta.
On the other hand, if the jet is Poynting dominated,
the emission radius may be large.
In any case, even non-detection can constrain the model parameters.

\section*{ACKNOWLEDGMENTS}

The authors would like to thank ~Shota~Kisaka,  Taka~Sakamoto, Masaomi~Tanaka, and Michitoshi~Yoshida
for helpful comments. 
This work is partly supported by
``New Developments in Astrophysics Through Multi-Messenger Observations of Gravitational
Wave Sources'', No. 24103006 (KI, TN), and
KAKENHI Nos. 15K05088 (R.Y.), 26287051, 26247042,
17H01126, 17H06131, 17H06362, 17H06357 (KI), No. 15H02087 (TN) by the Grant-in-Aid
from the Ministry of Education, Culture, Sports, Science and Technology (MEXT) of Japan.

\section*{Appendix~A: derivation of Eq.~(\ref{eq:jetthin})}
\label{sec:appe}

We assume that the emitting matter moves radially with Lorentz factor $\gamma$.
Then, the observed flux at the observed time $T$ and 
observed frequency $\nu$, measured in 
erg s$^{-1}$ cm$^{-2}$ Hz$^{-1}$, is given by
\begin{equation}
F_\nu(T)=\f{1}{D^2}\int^{2\pi}_0d\phi \int^1_{-1}d\mu
\int^{\infty}_0r^2dr\,
\f{j'_{\nu'} \left(\Omega_d', {\bm r}, T+{r\mu/c}\right)}
{\gamma^2(1- \beta \mu)^2}\ ,
\label{flux;j}
\end{equation}
where ${\bm r}=(r, \theta, \phi)$ and $\mu=\cos\theta$ \cite{Granot1999,Woods1999}.
Here, the quantities
 $\Omega_d'$ and $j'_{\nu'}$ 
are the direction towards the observer measured in the frame
comoving with the jet (comoving frame),
and the comoving frame emissivity 
in units of ergs s$^{-1}$ cm$^{-3}$ Hz$^{-1}$ sr$^{-1}$, respectively.
The frequency $\nu'$, which is measured in the comoving frame,
is given by $\nu'=\nu\gamma(1-\beta\mu)$.
The observer time $T$ and the time coordinate $t$ are related by
\begin{equation}
t=T+(r\mu/c)~~.
\label{eq:time_coordinates}
\end{equation}
%
The emissivity for the counter-jet emission has a functional form of
\beqa
j'_{\nu'}(\Omega_d', {\bm r}, t)&=&A(t) f(\nu')
\delta[r-r_0-\beta c(t-t_0)]
\nonumber\\
&\times& H(\Delta \theta-|\theta-\theta_v-\pi |)
H\left[\cos \phi-\left({{\cos\Delta \theta+\cos\theta
\cos\theta_v}\over
{-\sin \theta_v \sin\theta}}\right)\right],
\label{eq:emissivity_cjet}
\eeqa
where 
the Heaviside step function $H(x)$ describes that the emission is 
inside a cone of opening half-angle $\Delta \theta$.
Changing variables from $(r,\mu,\phi)$ to $(r,t,\phi)$ with Eq.~(\ref{eq:time_coordinates}),
we rewrite Eq.~(\ref{flux;j}) as
\begin{equation}
F_\nu(T)=\f{c}{D^2}\int dt\int^{2\pi}_0d\phi 
\int^{\infty}_0dr\, r\,
\f{j'_{\nu'} \left(\Omega_d', {\bm r}, t\right)}
{\gamma^2(1- \beta \mu)^2}\ ,
\label{flux;j2}
\end{equation}
After substituting Eq.~(\ref{eq:emissivity_cjet}) into Eq.~(\ref{flux;j2}),
one can first calculate the integration with respect to $r$ to derive Eq.~(\ref{eq:jetthin}).
Note that $\Delta\phi(t)$ is defined by
\begin{equation}
\Delta\phi(t)=\frac{1}{2}\int_0^{2\pi}d\phi\,
H\left[\cos \phi-\left({{\cos\Delta \theta+\cos\theta(t) 
\cos\theta_v}\over
{-\sin \theta_v \sin\theta(t)}}\right)\right]~~,
\end{equation}
which is rewritten as Eq.~(\ref{eq:delta_phi}).

\section*{Appendix~B: determining the model parameters}
\label{sec:appe2}

The starting time $T_{\rm start}^{\rm (f)}$ and ending time $T_{\rm end}^{\rm (f)}$ 
of the forward-jet emission are given by \cite{Yamazaki2003}
\begin{eqnarray}
T_{\rm start}^{\rm (f)}&=&T_0+({r_0}/{c\beta})
\left[1-\beta\cos(\max[0,\theta_v-\Delta\theta])\right],
\label{eq:tstart:xrf}
\\
T_{\rm end}^{\rm (f)}&=&T_0+[({r_0}/{c\beta})+t_e-t_0]
\left[1-\beta\cos(\theta_v+\Delta\theta)\right]~~.
\label{eq:tend:xrf}
\end{eqnarray}
Assuming $T_0=\tau_j$, $\beta\approx1$, $\theta_v < 1$, $\Delta \theta < 1$, and  $\theta_v>\Delta\theta$, we obtain
\begin{eqnarray}
T_{\rm start}^{\rm (f)}-\tau_j&=&\frac{r_0}{2\gamma^2c}
\left[1+\gamma^2(\theta_v-\Delta\theta)^2\right],
\label{eq:tstart_f}\\
T_{\rm end}^{\rm (f)}-\tau_j&=&\frac{\kappa r_0}{2\gamma^2c}
\left[1+\gamma^2(\theta_v+\Delta\theta)^2\right]~~.
\label{eq:tend_f}
\end{eqnarray}
One can solve Eqs.~(\ref{eq:tstart2}), (\ref{eq:tend2}), (\ref{eq:tstart_f}), and (\ref{eq:tend_f}) to have
$\gamma$, $\Delta\theta$, $r_0$, and $\kappa$
for given observed values of $\theta_v$, 
$T_{\rm start}^{\rm (c)}$, $T_{\rm end}^{\rm (c)}$, $T_{\rm start}^{\rm (f)}$, and $T_{\rm end}^{\rm (f)}$.
For that purpose, we define
\begin{eqnarray}
a&=&\frac{T_{\rm start}^{\rm (f)}-\tau_j}{T_{\rm start}^{\rm (c)}-\tau_j},\nonumber\\
b&=&\frac{T_{\rm end}^{\rm (f)}-\tau_j}{T_{\rm end}^{\rm (c)}-\tau_j}.
\end{eqnarray}
Then, we obtain from Eqs.~(\ref{eq:tstart2}) and (\ref{eq:tstart_f}),
\begin{equation}
a\gamma^2\left[4-(\theta_v+\Delta\theta)^2\right] =1+\gamma^2(\theta_v-\Delta\theta)^2~~.
\label{eq:a}
\end{equation}
Similarly,  from Eqs.~(\ref{eq:tend2}) and (\ref{eq:tend_f}), we get
\begin{equation}
b\gamma^2\left[4-(\theta_v-\Delta\theta)^2\right] =1+\gamma^2(\theta_v+\Delta\theta)^2~~.
\label{eq:b}
\end{equation}
Using Eqs.~(\ref{eq:a}) and (\ref{eq:b}), one can eliminate $\gamma^2$ to derive
 a quadratic equation for $\Delta\theta$:
 \begin{equation}
 (\Delta\theta)^2 +\frac{2(a+b-2)\theta_v}{a-b}\Delta\theta -(4-\theta_v^2)=0~~,
 \end{equation}
 which has a positive root because we consider the case of $\theta_v<1$.
Once  $\Delta\theta$ is known as a function of $\tau_j$, either Eq.~(\ref{eq:a}) or (\ref{eq:b}) gives us  $\gamma$ as a function of $\tau_j$.
Finally, with $\Delta\theta$,  $T_{\rm start}^{\rm (c)}$, $T_{\rm end}^{\rm (c)}$,
and $\theta_v$ from the GWs, 
we can determine 
$r_0$ and $\kappa$ by solving Eqs.~(\ref{eq:tstart2}) and (\ref{eq:tend2})  as a function of $\tau_j$.
In summary, we can estimate $\Delta\theta$, $\gamma$, $r_0$ and $\kappa$
as functions of $\tau_j$,
which may be determined by the method in Appendix~C.

\section*{Appendix~C: Forward and counter jets for $\theta_v\sim \pi/2$}
\label{sec:appe3}
Let us consider the case of $\theta_v=\pi/2-\delta\theta_v$ with $0\leq\delta\theta_v < 1$ . This means that the line of sight to the sGRB is almost
perpendicular to the jet direction. Then, from 
Eqs.~(\ref{eq:tstart:uvf}), (\ref{eq:tend:uvf}),
(\ref{eq:tstart:xrf}), and (\ref{eq:tend:xrf}), with $c\beta t_0=r_0$ and $c\beta t_e=r_e$, we have
\begin{eqnarray}
T_{\rm start}^{\rm (f)}-\tau_j&=&\frac{r_0}{c\beta}\left[1-\beta\sin(\delta\theta_v+\Delta\theta)\right],\\
T_{\rm end}^{\rm (f)}-\tau_j&=&\frac{r_e}{c\beta}\left[1-\beta\sin(\delta\theta_v-\Delta\theta)\right],\\
T_{\rm start}^{\rm (c)}-\tau_j&=&\frac{r_0}{c\beta}\left[1+\beta\sin(\delta\theta_v-\Delta\theta)\right],\\
T_{\rm end}^{\rm (c)}-\tau_j&=&\frac{r_e}{c\beta}\left[1+\beta\sin(\delta\theta_v+\Delta\theta)\right].
\end{eqnarray}
Since $\beta\sim 1$ and $\sin(\delta\theta_v\pm\Delta\theta)\sim \delta\theta_v\pm\Delta\theta$, we get
\begin{equation}
\frac{T_{\rm start}^{\rm (f)}-\tau_j}{T_{\rm start}^{\rm (c)}-\tau_j}=\frac{T_{\rm end}^{\rm (f)}-\tau_j}{T_{\rm end}^{\rm (c)}-\tau_j},
\end{equation}
within the first order of perturbations.
This equation is solved for $\tau_j$ as
\begin{equation}
\tau_j = \frac{T_{\rm end}^{\rm (c)}T_{\rm start}^{\rm (f)}-T_{\rm start}^{\rm (c)}T_{\rm end}^{\rm (f)}}
{(T_{\rm end}^{\rm (c)}-T_{\rm start}^{\rm (c)})-(T_{\rm end}^{\rm (f)}-T_{\rm start}^{\rm (f)})} ~.
\end{equation}
Therefore, from the observed values of $T_{\rm start}^{\rm (f)}, T_{\rm start}^{\rm (c)}, T_{\rm end}^{\rm (f)}$ and $T_{\rm end}^{\rm (c)}$,
we can determine $\tau_j$.



\end{document}